\documentstyle[editedbook,epsfig,psfig,epsf]{mq}
\def\eg{{\it e.g.} }
\def\etal{{\em et al.} }

\def\cm2{cm$^2$ }
\def\se1{s$^{-1}$ }

\begin{opening}
\title{The Hard Truth about Some ``Soft'' X-ray Transients}
\author{R. M. Bandyopadhyay$^1$, C. Brocksopp$^{2}$, \& R. P. Fender$^3$}
\institute{$^1$ Dept. of Astrophysics, University of Oxford, Keble Road, Oxford, OX1 3RH  U.K.\\
$^2$ Astrophysical Research Institute, Liverpool John Moores University, Twelve Quays House, Egerton Wharf, Birkenhead CH41 1LD U.K.\\
$^3$ Astronomical Institute ``Anton Pannekoek'' and Center for High-Energy Astrophysics, University of Amsterdam, Kruislaan 403, 1098 SJ Amsterdam, The Netherlands}
\end{opening}

\runningtitle{The Hard Truth about Some ``Soft'' X-ray Transients}
\runningauthor{Bandyopadhyay, Brocksopp, \& Fender}

\begin{document}
\vspace{-0.5cm}
\begin{abstract}
{\small We have accumulated multiwavelength lightcurves for eight
black hole X-ray binaries which have been observed to enter a supposed
``soft X-ray transient'' outburst, but which in fact remained in the
low/hard state throughout the outburst.  Comparison of the lightcurve
morphologies, spectral behaviour, properties of the QPOs and the radio
jet provides the first study of such objects as a subclass of X-ray
transients (XRTs).  However, rather than assuming that these hard
state XRTs are different from ``canonical'' soft XRTs, we prefer to
consider the possibility that a new analysis of both soft and hard
state XRTs in a spectral context will provide a model capable of
explaining the outburst mechanisms for the majority of black hole
X-ray binaries.}
\end{abstract}

\section{Lightcurves}
In the low/hard state (LHS) the X-ray spectrum is dominated by a power
law component; this hard X-ray emission is thought to be produced in a
Comptonizing corona.  Corresponding X-ray power spectra for sources in
the LHS show a high level of low frequency noise, a broken power law,
and at least one quasi-periodic oscillation (QPO).  The LHS is also
characterized by a powerful, low intensity jet emitting synchrotron
radiation in radio (and often higher) frequencies \cite{Fen01}.

The X-ray lightcurves of the eight LHS XRTs we consider -- V404~Cyg,
A~1524-62, 4U~1543-475, GRO~J0422+32, GRO~J1719-24, GRS~1737-31,
GS~1354-64, and XTE~J1118+480 -- exhibit very different morphologies,
and the canonical ``FRED''-type lightcurve is not predominant.  The
optical lightcurves appear generally, although not consistently,
correlated with the X-rays.  Where radio coverage is available it
appears that both the main outburst and secondary maxima of the X-ray
lightcurves tend to be associated with radio ejections \cite{Fen01}.
Despite the similarities in X-ray and broad-band spectral behaviour of
these sources, the most notable feature of these lightcurves is their
inconsistency.

\section{The Low-Frequency QPO}

\begin{figure}[htb]
\centering 
\epsfig{file=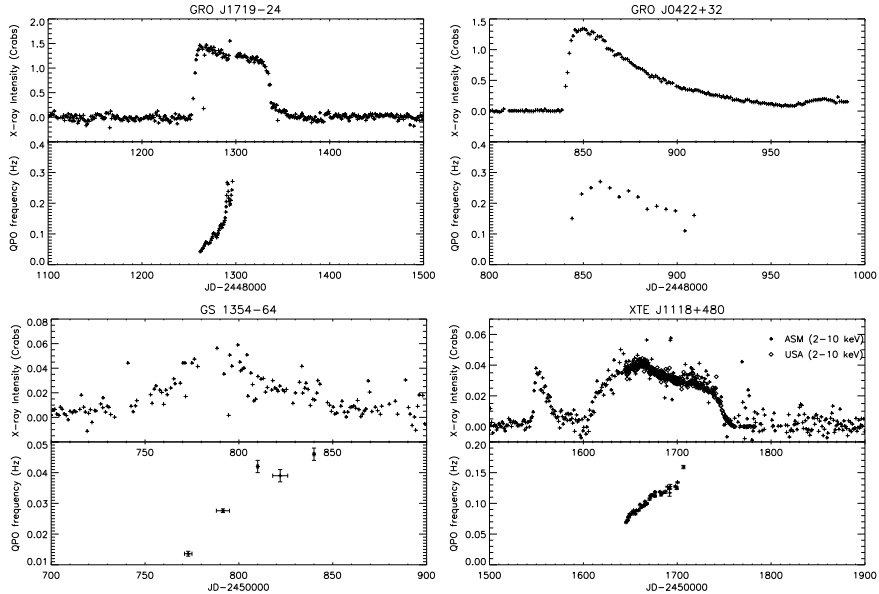,width=12cm}
\caption{The variability of the low frequency QPO in four low/hard transient outbursts.  The increase in QPO frequency is generally not well correlated with the X-ray lightcurve.}
\label{fig:fig1}
\vspace{-0.4cm}
\end{figure}

The presence of a QPO in the power spectrum of XRTs is a common
feature of the LHS, as well as the intermediate and very high states
(I/VHS; \cite{WvdK99}); however, the frequencies are lower in the LHS
than in the I/VHS.  In the four sources shown in Figure 1, the
frequency of the LHS QPO is not constant but instead increases during
the outburst.  Formal cross-correlation of the QPOs with the X-ray
flux does not produce a statistically significant correlation in any
of these sources.  In addition, we find that the duration of the QPO
frequency increase is a factor of 1.5-2 times the duration of the
X-ray flux rise.

The mHz QPO of XRTs may relate to the inner edge of the accretion disc
\cite{Rev00}; if the softening during the outburst is due to the inner
edge of the disc moving inwards we would expect the QPO frequency to
increase, as observed.  Alternatively, the QPO could be produced at a
large radius within the disc, with its frequency inversely
proportional to the disc mass \cite{Wood01}.  In this model, the
rising QPO frequency is a signature of the accretion disc mass
decreasing as it disappears into the black hole.

\section{Broad-band Spectra and Jets}

\begin{figure}[htb]
\centering 
\epsfig{file=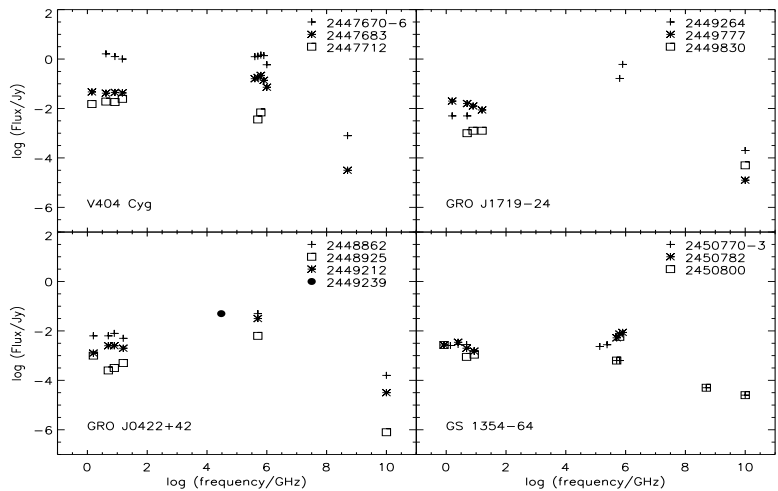,angle=90,height=6.4cm,width=10.2cm}
\caption{Broad-band spectra for four low/hard transient outbursts.  Three epochs are plotted for each, corresponding (except in GS 1354-64) to the outburst peak (crosses), secondary peak (asterisks), and periods of low luminosity (squares).}
\label{fig:fig2}
\vspace{-0.4cm}
\end{figure}

Nearly all black hole transients in the LHS show evidence for some
form of jet behaviour \cite{Fen01}.  This jet is not purely a ``radio
jet'', as the synchrotron spectrum that is thought to be the jet
signature is often seen up to IR and possibly optical frequencies.
For XTE~J1118+480, a synchrotron spectrum has been fit to the full
range of frequencies up to the hard X-rays \cite{Sera01}.  For these
LHS XRTs, a flat or inverted spectrum has already been well
established in the radio.  We compiled spectra from radio to X-rays
for four sources at three different epochs (where possible) per source
(Figure 2).  Inspection of the resultant broad-band spectra show that
despite the morphological differences between the outburst
lightcurves, their spectral properties are similar; also, the spectra
do not vary significantly between the various epochs.  The spectra of
these four XRTs are similar to that of XTE~J1118+480; formal fitting
of these broad-band spectra with a synchrotron spectrum is in
progress.  Although it is unclear whether synchrotron emission from
the jet can be the dominant contributor to the broad-band spectrum, as
suggested in the case of XTE~J1118+480, what is clear is that the jet
contribution should not be ignored when modelling XRT outbursts,
especially in the LHS.

\section{Conclusions}
For these LHS transient outbursts, we find the following characteristics:
\begin{itemize}
\item{The X-ray and multi-wavelength lightcurves have very different
morphologies.  The relationships between the emission at various
wavelengths likewise differs from source to source.}
\item{A low frequency QPO is observed which increases in frequency
during the outburst but is not directly correlated with the X-ray
luminosity.}
\item{The broad-band spectra of the LHS transients are very similar
and do not vary substantially during different epochs.}
\item{In addition to the radio signature of the jet, it may be
possible to fit the broad-band spectra in the LHS with a synchrotron
spectrum.}
\end{itemize}

To model LHS outbursts, the Disc Instability Model (DIM) should be
able to reproduce non-FRED lightcurves and also must consider the
production of the power law hard X-ray emission from the corona.
Models of LHS outbursts should include the jet, which is likely to be
a much more significant contributor to the X-ray luminosity and
broad-band emission than has been previously assumed.  Finally, in
recent ``canonical'' soft XRT outbursts (\eg XTE~J1859+226,
XTE~J1550-564) the sources have been observed to pass through the LHS
on the rise from quiescence to the (V)HS.  It is important that the
power requirements of the initial LHS and its associated jet are
incorporated into future attempts to model transient outbursts with
the DIM, which has not been done to date.

\end{document}